\documentclass[10pt,russian,english]{article}
\usepackage{amssymb}
    \usepackage{babel}
\usepackage[T2A]{fontenc}
\usepackage{graphicx}

\begin{document}
\title{Superfluidity of electron-hole pairs in randomly inhomogeneous bilayer systems}

\author{A. I. Bezuglyj$^1$, S. I. Shevchenko$^2$}

\date{}

\maketitle

\noindent

$^1${\it NSC Kharkov Institute of Physics and Technology, 61108, Kharkov, Ukraine}

 \vskip .5 cm

$^2${\it B.\,I.\, Verkin Institute for Low Temperature Physics and Engineering, 61103, Kharkov, Ukraine }

E-mail: shevchenko@ilt.kharkov.ua

 \vskip 1.5 cm

\baselineskip .5 cm

\begin{abstract}
In bilayer systems electron-hole ($e-h$) pairs with spatially separated components (i.e., with electrons
in one layer and holes in the other) can be condensed to a superfluid
state when the temperature is lowered. This article deals with the influence of randomly distributed inhomogeneities on the
superfluid properties of such bilayer systems in a strong perpendicular magnetic field. Ionized
impurities and roughenings of the conducting layers are shown to decrease the superfluid current density of the $e-h$
pairs. When the interlayer distance is smaller than or close to the magnetic length,  the  fluctuations of the
interlayer distance considerably reduce the superfluid transition temperature.
\end{abstract}

\vskip 1 cm


\vskip 1 cm

\textheight 22.5 cm

\topmargin=-0.5in

%\pagebreak
\section{Introduction}

The bilayer systems with pairing of electrons of one layer with holes of the other layer have attracted
considerable interest \cite{1,2,3} because such electron-hole ($e-h$) pairs can be condensed to a superfluid state,
accompanied by a counterflow superconductivity (as predicted in Refs. 4,5). This special type of
superconductivity manifests itself as equal and antiparallel nondissipative currents flowing in the conducting
layers when the $e-h$ pairs move in one direction. Experiments on antiparallel currents were carried out with GaAs/AlGaAs
heterostructures formed by two closely spaced quantum wells, each containing a quasi-two-dimensional electron
gas \cite{6,7,8}. It has been found that in a perpendicular magnetic field, when the total Landau level
filling factor $\nu_T$ is equal to unity ($\nu_T=\nu_1+\nu_2 = 1$), a lowering of the temperature leads to a
significant (more than an order of magnitude) decrease of the longitudinal resistance $\rho_{xx}$ and the Hall
resistance $\rho_{xy}$ in each layer. The obtained results can be interpreted in terms of superfluidity of the $e-h$ pairs with
spatially separated components (interwell excitons) \cite{7,9}. Indeed, the condition $\nu_T = 1$ is equivalent
to the equality $\nu_1 = 1 -\nu_2$ which means that the number of occupied states in the first layer exactly
coincides with the number of vacant states (holes) in the second layer. If the temperature $T=0$, all electrons
from the first layer and all holes from the second layer form $e-h$ pairs and there are no unpaired carriers.
Due to superfluidity of the $e-h$ pairs the longitudinal resistance $\rho_{xx}$ in each layer is equal to zero. The
Hall resistance $\rho_{xy}$ must also become zero because $e-h$ pairs (being electrically neutral) does not
contribute to the Hall effect. The residual values of $\rho_{xx}$ and $\rho_{xy}$ observed in the experiment (at
$T\neq 0)$ are caused by motion of free vortices and unpaired carriers respectively.

The properties of the superfluid  state of $e-h$ pairs in the bilayer system  depend  critically on the
interlayer distance $d$. Experiments show that the decrease of $\rho_{xx}$ and $\rho_{xy}$ described above is
not observed if $d$ exceeds $1.6 l_H$ - $1.9 l_H$, where $l_H=\sqrt{\hbar c/eH}$ is the magnetic length
\cite{6,8}. According to theoretical concepts \cite{10}, with increasing $d$ the coherence of $e-h$ pairs breaks
down if the coherence length becomes smaller than the average distance between the pairs. This transition is of a
quantum nature, since at large $d$ the interlayer coherence is destroyed by quantum fluctuations \cite{11}.

The main attention in this article is paid to the properties of systems with relatively small interlayer
distances $d \lesssim l_H$. This range of interlayer distances is quite important because in a homogeneous
system the critical temperature of the superfluid transition ($T_c$) reaches its maximum value at $d \approx 0.4
l_H$ (see later). The maximum is rather sharp, since with reducing $d$ the interaction between the $e-h$ pairs is
weakened, and the value of $T_c$ decreases rapidly due to thermal fluctuations. It turns out that in an
inhomogeneous bilayer system with small $d$ the weakness of the interaction between the $e-h$ pairs leads to a high sensitivity
of $T_c$ to the structure inhomogeneities. As a consequence, the critical temperature is strongly lowered by
inhomogeneities for $d \lesssim l_H$. (See Ref. \cite{12} and the references cited therein for the influence
of inhomogeneities on the bilayer system at $d \gg l_H$.)

%\pagebreak

\section{Suppression of superfluidity of $e-h$ pairs by the electric field of ionized donors}

In the bilayer systems that are based on  semiconductor heterostructures, electrons fill of the quantum wells
due to ionization of donor atoms. We will show that the electric field of the donor layer decreases the
superfluid current of $e-h$ pairs in the bilayer system. This decrease is caused by correlations between spatial
fluctuations of superfluid density $n_s$ and superfluid velocity ${\bf v}_s$, namely, under influence of the
donor electric field the superfluid density $n_s$ becomes inhomogeneous, and due to the current conservation law
${\rm div}\,{\bf j}_s={\rm div}(n_s{\bf v}_s) =0$ in the regions where $n_s$ is higher the value of $v_s$ is
lower. These correlations lead to decreasing of the average superfluid current density in the second order of
the perturbation theory.

The following analysis is based on the nonlinear equation for the wave function of the $e-h$ pair condensate
obtained in Ref. 10:

\begin{eqnarray}\label{1}
i\hbar\frac{\partial\Psi({\bf R},t)}{\partial t} = \Bigl\{ \Bigl(i\hbar \frac{\partial}{\partial{\bf
R}}-{\alpha({\bf R})\over c}\Bigl[{\bf E}_{tot}{\bf H}\Bigr]\Bigr){1\over 2M_H ({\bf R})} \Bigl(i\hbar
\frac{\partial}{\partial{\bf R}}-{\alpha({\bf R})\over c}\Bigl[{\bf E}_{tot}{\bf H}\Bigr]\Bigr) \nonumber \\
 - {1\over 2}\alpha({\bf R}) E_{tot}^2 + U({\bf R})-\mu +g({\bf R})|\Psi({\bf R},t)|^2 \Bigr\} \Psi({\bf R},t).
\end{eqnarray}

Note that the equation (\ref{1}) is obtained within the self-consistent field approximation at low density of
$e-h$ pairs ($\nu_1 \ll 1 $). The equation (\ref{1}) is similar to the Gross - Pitaevskii equation \cite{13},
that is not surprising since at low density the $e-h$ pairs act as a weakly nonideal Bose gas \cite{14,15}. Due
to inhomogeneity of the bilayer system the coefficients in (\ref{1}) (the effective mass of the pair $M_H$, its
polarizability $\alpha$ and the interaction constant $g$) are functions of coordinates in the plane of the
layers: ${\bf R}=(x,y)$. Explicit expressions for these coefficients and for the potential energy of the $e-h$
pairs $U({\bf R})$ will be shown below. The significant difference of (\ref{1}) from the Gross - Pitaevskii
equation consists in changing the structure of the first term in the right hand side. Firstly, the action of the
gradient operator spreads to the coordinate dependent mass of the $e-h$ pair, and secondly, the gradient
operator itself is "extended" with an addition related to the polarizability of the $e-h$ pairs. This addition
contains the effective electric field ${\bf E}_{tot}={\bf E}_{\|}+{\bf E}_{int}$, where ${\bf E}_{\|}$ is the
electric field of the external charges (parallel to the layers) and ${\bf E}_{int}$ is the field caused by
curvature of the conducting layers \cite{10,16}.

It is easy to demonstrate that the imaginary part of the equation (\ref{1}) is
reduced to the continuity equation $\dot n_s + {\rm div}{\bf j}_s = 0$, where
the superfluid component density is $n_s=\mid \Psi \mid ^2$ and the superfluid
current density is

\begin{eqnarray}\label{2}
{\bf j}_s=\frac{n_s}{M_H} \Bigl(\hbar \frac{\partial \varphi}{\partial{\bf R}}+{\alpha({\bf R})\over
c}\Bigl[{\bf E}_{tot}{\bf H}\Bigr]\Bigr).
\end{eqnarray}
Here $\varphi$ is the phase of the wave function ($\Psi =\mid \Psi \mid e^{i\varphi}$).

\begin{figure}
  \includegraphics[width=9 cm]{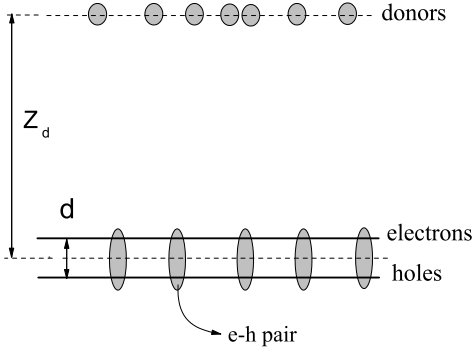}
  \caption{Schematic view of a bilayer electron-hole system with interlayer distance $d$. Electron-hole pairs with
spatially separated components and a layer of donor atoms located at a distance $Z_d$ from the bilayer system
are shown.}\label{f1}
\end{figure}

Now we consider the case when the inhomogeneity is caused by random positions of ionized donor atoms in a layer
at a distance of $Z_d$ from the bilayer system (Fig. \ref{f1}). In such a situation the coefficients in Eq.
(\ref{1}) do not depend on coordinates and ${\bf E}_{tot}$ coincides with the parallel electric field ${\bf
E}_{\|}$. The field ${\bf E}_{\|}$ and the potential energy of an $e-h$ pair $U({\bf R})$ are expressed in terms
of the electrostatic potentials $V_{1}({\bf R})$ and $V_{2}({\bf R})$ created by the charged donors in the
electron and hole layers correspondingly: ${\bf E}_{\|}= -(1/2)\frac{\partial }{\partial{\bf R}}[V_{1}({\bf
R})+V_{2}({\bf R})]$; $U({\bf R})=e[V_{2}({\bf R})-V_{1}({\bf R})]$. In this case the real part of the equation
(\ref{1}) can be written in the form

\begin{equation}\label{3}
\frac{\hbar^2}{2 M_H}\frac{1}{\sqrt{n_s}}\frac{\partial^2 \sqrt{n_s}}
 {\partial{\bf R}^2}-\frac{M_H}{2}v_s^2+\frac{\alpha}{2 }
E_{\parallel}({\bf R})^2+e[V_{2}({\bf R})-V_{1}({\bf R})]-g n_s+\mu = 0.
\end{equation}

Note that experiments with bilayer systems are carried out in strong magnetic
fields, when the magnetic length is much less than the Bohr radius of the
$e-h$ pair. In this limit
the effective mass of the $e-h$ pair $M_H$ appears as the result of Coulomb
interaction between the electron and the hole \cite{17,18} and its value does
not depend on the intrinsic masses of the carriers:

\begin{equation}\label{4}
M_H = {2\varepsilon\hbar^2\over e^2l_H}\sqrt{{2\over\pi}} \Bigl[\Bigl(1+{d^2\over
l_H^2}\Bigr)\exp\Bigl({d^2\over 2l_H^2}\Bigr) {\rm erfc}\Bigl({d\over \sqrt2 l_H}\Bigr)-\sqrt{{2\over\pi}}
{d\over l_H}\Bigr]^{-1}.
\end{equation}
Here $\varepsilon $ is the dielectric constant and ${\rm erfc}(x)=1-\frac{2}{\sqrt \pi} \int_0^x e^{-t^2} dt$ is
the complementary error function. Besides $M_H$, the equation (\ref{3}) also contains the polarizability of the
$e-h$ pair $\alpha=c^2 M_H/H^2$ and the interaction constant

\begin{equation}\label{5}
g={e^2 \over \varepsilon}\Bigl[4\pi d - (2\pi)^{3/2}l_H +(2\pi)^{3/2}l_H \exp\Bigl({d^2\over 2l_H^2}\Bigr) {\rm
erfc}\Bigl({d\over \sqrt2 l_H}\Bigr)\Bigr].
\end{equation}

The expression for the electrostatic potentials created by donors in the
electron and hole layers has the form

\begin{equation}\label{6}
V_{1(2)}={e \over \varepsilon}\int d{\bf R}' \rho ({\bf R}')\Bigl[({\bf R}-{\bf R}')^2 +  \Bigl(Z_d \mp {d\over
2 }\Bigr)^2\Bigr]^{-\frac {1}{2}},
\end{equation}
where the density of donor atoms is $\rho ({\bf R}')=\sum_{i} \delta({\bf R}'-{\bf R}_i)$. In this equality
${\bf R}_i$ determines the position of the $i$-th donor.

If the potentials $V_{1(2)}({\bf R})$ change significantly at distances much larger than the coherence length
$\xi = \hbar/\sqrt{M_H g n_s}$ then the first term in (\ref{3}) can be neglected as small relatively to $gn_s$.
Since  we are interested in the linear in $V_{1(2)}$ correction to $n_s$, we also omit the second and third
terms which are quadratic in $n_s$. As a result we obtain the equation for $n_s$ in the Thomas -- Fermi
approximation:

\begin{equation}\label{7}
e[V_{2}({\bf R})-V_{1}({\bf R})]-g n_s({\bf R})+\mu = 0.
\end{equation}

It is convenient to split the potential difference $V_{1}({\bf R})-V_{2}({\bf R})$ to the homogeneous part
$\langle V_{1}-V_{2}\rangle=\frac{1}{S} \int d{\bf R}[V_{1}({\bf R})-V_{2}({\bf R})]$, where $S$ is the system
area, and the deviations from the average value $\tilde V_{1}-\tilde V_{2}= V_{1}-V_{2}-\langle
V_{1}-V_{2}\rangle$. The homogeneous part determines the average filling factors $\nu_1$ and $\nu_2$, whereas
the inhomogeneous part gives the spatial fluctuations of the superfluid density of $e-h$ pairs we are interested
in.

\begin{equation}\label{8}
\tilde n_s ({\bf R}) = \frac{e}{g} [\tilde V_{1}({\bf R})-\tilde V_{2}({\bf R})].
\end{equation}
Since $n_s$ and ${\bf v}_s$ are related by the equation ${\rm div}(n_s{\bf v}_s) =0$, the density fluctuations
$\tilde n_s$ lead to fluctuations of the superfluid velocity $\tilde {\bf v}_s$. In the linear approximation we
have

\begin{equation}\label{9}
 {\bf v}_{s0} \frac{\partial \tilde n_s }{\partial{\bf R}}+ n_{s0} \, {\rm div} \, \tilde {\bf v}_s =0,
\end{equation}
where the index "0" denotes the homogeneous part. This scalar equation is insufficient to find $\tilde {\bf
v}_s$. An additional equation can be obtained from (\ref{2}). In the absence of quantized vortices

\begin{equation}\label{10}
{\rm curl} \, \tilde {\bf v}_s =- \frac{\alpha {\bf H} }{c M_H}{\rm div} \, \tilde {\bf E}_\|.
\end{equation}
In order to find $\tilde {\bf v}_s$ we calculate curls of both sides of this equality. For the Fourier component
of the superfluid velocity $v_{s \bf k}$ we obtain
\begin{equation}\label{11}
k^2 {\bf v}_{s {\bf k}}- {\bf k}({\bf k} {\bf v}_{s{\bf k}}) = \frac{\alpha }{c M_H}({\bf k} {\bf E}_{\|{\bf
k}})[{\bf k }{\bf H }].
\end{equation}
Equation (\ref{9}) leads to $({\bf k} {\bf v}_{s{\bf k}})= -({\bf k} {\bf v}_{s0}) n_{s{\bf k} }/ n_{s0} $.
Substituting this expression into (\ref{11}) yields

\begin{equation}\label{12}
{\bf v}_{s {\bf k}}=- \frac {n_{s{\bf k} }}{ n_{s0}} \frac {{\bf k}({\bf k} {\bf v}_{s0})}{k^2} + \frac{\alpha
}{c M_H} \frac {({\bf k} {\bf E}_{\|{\bf k}})}{k^2}[{\bf k }{\bf H }].
\end{equation}
Thus, fluctuations of the superfluid velocity $\tilde {\bf v}_s$ are determined
by fluctuations of the superfluid density $n_s$ and fluctuations of the
electric field ${\bf E}_{\|}$.

Since in real bilayer systems $Z_d \gg d$, when calculating the Fourier
component of the parallel electric field ${\bf E}_{\|{\bf k}}$ we may assume
$d=0$. Using the definition of ${\bf E}_{\|{\bf k}}$, we obtain from (\ref{6}):

\begin{equation}\label{13}
{\bf E}_{\|{\bf k}}= - \frac {2\pi i e }{\varepsilon}\rho_{{\bf k}} e^{-k Z_d} \frac{{\bf k}}{k},
\end{equation}
where $\rho_{{\bf k}}=\sum_i  e^{i {\bf k}{\bf R}_i}$. If we expand $V_{1(2)}$
by $d$, the Fourier component of the superfluid density can be easily found
from (\ref{8}):

\begin{equation}\label{14}
n_{s{\bf k}}=  \frac {2\pi d e^2 }{\varepsilon g}\rho_{{\bf k}} e^{-k Z_d}.
\end{equation}

After substituting the formulae obtained above into the expression for the average superfluid current density
$\langle{\bf j}_s\rangle=\langle n_s{\bf v}_s\rangle = n_{s0}{\bf v}_{s0} + \sum_{{\bf k}\neq 0} n_{s{\bf k}}
{\bf v}_{s \,-{\bf k}}$ we obtain

$$\langle{\bf j}_s\rangle = n_{s0}{\bf v}_{s0} - \frac {1}{n_{s0}} \sum_{{\bf k}\neq 0}\Bigl(\frac {2\pi d e^2
}{\varepsilon g}\Bigr)^2 \rho_{{\bf k}} \rho_{-{\bf k}} e^{-2 k Z_d}\frac {{\bf k}({\bf k} {\bf v}_{s0})}{k^2}
+$$
\begin{equation}\label{15}
i\sum_{{\bf k}\neq 0}  \frac {4{\pi}^2 e^2 d \alpha}{\varepsilon^2 g c M_H} \rho_{{\bf k}} \rho_{-{\bf k}}e^{-2
k Z_d}\frac{[{\bf k }{\bf H }]}{k}.
\end{equation}
Here and below it is assumed that the area of the system $S=1$. After averaging (\ref{15}) over positions of the
donor atoms (which is reduced to replacing $\rho_{{\bf k}} \rho_{-{\bf k}}$ with the donor density $n_d$) and
changing summation by ${\bf k}$ to integration, we note that the second integral becomes zero due to integration
over angles of the vector ${\bf k}$. Calculating of the remaining integral does not present difficulties and
yields

\begin{equation}\label{16}
\langle{\bf j}_s\rangle = \Bigl[ n_{s0}-\frac{\pi}{4}\Bigl(\frac { d e^2 }{\varepsilon g}\Bigr)^2\frac{n_d}{
n_{s0}}\frac{1}{Z_d^2}\Bigr]{\bf v}_{s0}.
\end{equation}

The expression in square brackets is the effective superfluid density of $e-h$ pairs, $n_{ef}$. As follows from
(\ref{16}), $n_{ef} < n_{s0}$, thus, under the action of the electric field of randomly positioned donors the
average density of superfluid current decreases.

Now we consider in detail the case of small interlayer distances. At $d \ll
l_H$ the interaction constant of the $e-h$ pairs

\begin{equation}\label{17}
g=\sqrt{2} \pi^{3/2} {e^2 d^2\over \varepsilon l_H},
\end{equation}
and for the effective density of the superfluid component we find an expression

\begin{equation}\label{18}
n_{ef}  = n_{s0} - \frac{1}{8\pi^2 Z_d^2}\Bigl(\frac{n_d}{ n_{s0}}\Bigr)\Bigl(\frac{l_H^2}{ d^2}\Bigr).
\end{equation}

Since $n_{d}/n_{s0}\sim 1/\nu_1 \gg 1$ and $l_H \gg d$, in this case a superfluid state exists only at quite
large distance of donor layer from the bilayer system. Another aspect of the obtained result is that at given
$Z_d$ the destructive influence of donors can be reduced by decreasing $l_H $ to values about $d$ and increasing
$\nu_1$ to the maximum value 1/2. Note that just at $d \sim l_H$ and $\nu_1 = \nu_2 = 1/2$ bilayer systems
possess properties which can be explained by appearance of superfluidity of $e-h$ pairs \cite{6,7,8}.

Since two layers of donor atoms are used in experiments, one of which is
locates above the bilayer system and another is under it, the expression
(\ref{16}) must be generalized to this case. Calculations (analogous to that
for one layer) yield

\begin{equation}\label{19}
\langle{\bf j}_s\rangle = \Bigl[ n_{s0}-\frac{\pi}{4}\Bigl(\frac { d e^2 }{\varepsilon
g}\Bigr)^2\Bigl(\frac{n_{d1}}{ Z_{d1}^2}+\frac{n_{d2}}{ Z_{d2}^2}\Bigr)\frac{1}{ n_{s0}}\Bigr]{\bf v}_{s0},
\end{equation}
where $n_{di}$ is the density of donor atoms in the i-th layer, $Z_{di}$ is the
distance from the bilayer system to the i-th donor layer (i = 1,2).

%\pagebreak

\section{Suppression of superfluidity of $e-h$ pairs by variations of interlayer distance}

In this section we will show that in bilayer systems with curved layers variations of the interlayer distance
lead to a decrease of the average superfluid current density. So, let the interlayer distance be a smooth
function of coordinates in the plane of the layers: $d = d({\bf R})$. Since the effective mass of a pair $M_H$,
its polarizability $\alpha$ and the interaction constant $g$ depend on $d$, it turns out that they also depend
on ${\bf R}$. In the absence of electric fields created by external charges the potential energy of the $e-h$
pair coincides with its binding energy which, according to \cite{10}, has the form

\begin{equation}\label{20}
E_0 = - {e^2 \over \varepsilon l_H }\sqrt{\pi \over 2} \exp\Bigl({d^2\over 2l_H^2}\Bigr) {\rm erfc}\Bigl({d\over
\sqrt2 l_H}\Bigr).
\end{equation}
Owing to the dependence of $E_0$ on $d$ a random potential profile occurs in the inhomogeneous system: $U({\bf
R}) = E_0 [d({\bf R})]$. Finally we remind that curvature of the conducting layers leads to a field ${\bf
E}_{int}$ which polarizes the $e-h$ pair \cite{10}:

\begin{equation}\label{21}
{\bf E}_{int} =  {e \over 2 \varepsilon l_H } F(d) \frac{\partial }{\partial{\bf R}}[z_{1}({\bf R})+z_{2}({\bf
R})].
\end{equation}
Here $z_{1(2)}({\bf R})$ denotes the $z$ coordinate of the electron (hole)
layer, and the dimensionless function $F(d)$ is determined by the expression

\begin{equation}\label{22}
F(d) = - {d^2\over 2l_H^2} \Bigl[ 1 - \sqrt{2 \pi} {d\over \ l_H}\exp\Bigl({d^2\over 2l_H^2}\Bigr) {\rm
erfc}\Bigl({d\over \sqrt2 l_H}\Bigr)\Bigr].
\end{equation}

In the inhomogeneous bilayer system where the effective mass of a pair is a function of coordinates it is
convenient to write down the expression for the superfluid current density as ${\bf j}_s({\bf R})=K({\bf R}){\bf
p}_s({\bf R})$. Here the coefficient

\begin{equation}\label{23}
K({\bf R}) =  {n_s({\bf R})\over M_H({\bf R})},
\end{equation}
and the kinematic momentum of the $e-h$ pair

\begin{equation}\label{24}
{\bf p}_s({\bf R}) = \hbar \frac{\partial \varphi}{\partial{\bf R}}+{\alpha({\bf R})\over c}\Bigl[{\bf
E}_{int}{\bf H}\Bigr].
\end{equation}

To find $\langle {\bf j}_s({\bf R})\rangle$ we perform calculations analogous
to the calculations in the previous section. $z_{1(2)}({\bf R})$ can be
expressed as a sum of the average value and the fluctuation addition:
$z_{1(2)}({\bf R})= \langle z_{1(2)}\rangle + \tilde z_{1(2)}({\bf R})$. In the
linear order in fluctuations the current conservation condition ${\rm div}{\bf
j}_s = 0$ gives a relation between Fourier components $K_{\bf k}$ and ${\bf
p}_{s{\bf k}}$:

\begin{equation}\label{25}
({\bf k} {\bf p}_{s{\bf k}})= -({\bf k} {\bf p}_{s0}) K_{{\bf k} }/K_0.
\end{equation}
Calculating now ${\rm curl \,curl}\,{\bf p}_{s}$ and using (\ref{25}), we obtain

\begin{equation}\label{26}
 {\bf p}_{s{\bf k}}= -\frac{K_{\bf k}}{K_0}
 \frac{({\bf k} {\bf p}_{s0})}{k^2} + {\alpha \over c}\Bigl[{\bf E}_{int \,{\bf k}}{\bf H}\Bigr],
\end{equation}
where the polarizability $\alpha $ is taken in the zeroth order in fluctuations. After substituting ${\bf
p}_{s\, -{\bf k}}$ into the expression for the average superfluid current $\langle{\bf j}_s\rangle = K_{0}{\bf
p}_{s0} + \sum_{{\bf k}\neq 0} K_{\bf k} {\bf p}_{s\, -{\bf k}}$, we have

\begin{equation}\label{27}
\langle{\bf j}_s\rangle =  K_{0}{\bf p}_{s0} -\frac{1}{K_0} \sum_{{\bf k}\neq 0} K_{\bf k}K_{-{\bf k}}
 \frac{{\bf k} ({\bf k} {\bf p}_{s0})}{k^2} + {\alpha \over c}\sum_{{\bf k}\neq 0} K_{\bf k}
 \Bigl[{\bf E}_{int \,-{\bf k}}{\bf H}\Bigr].
\end{equation}

To find the superfluid density $n_s$ appearing in the coefficient $K$ (\ref{23}), one must use the equation
(\ref{1}). For the case considered the Thomas - Fermi approximation reduces the real part of (\ref{1}) to the
equality

\begin{equation}\label{28}
E_0 [d({\bf R})] + g({\bf R}) n_s({\bf R}) - \mu = 0.
\end{equation}
Equation (\ref{28}) can be split to the homogeneous and fluctuating parts. The homogeneous part determines
$n_{s0}$, and the fluctuating part gives $\tilde n_s$ -- deviations of the superfluid density from to the
average value. In the linear approximation in fluctuations

\begin{equation}\label{29}
\tilde n_s ({\bf R}) = - \frac{1}{g}\Bigl[\frac{\partial E_0}{\partial d} + n_{s0}\frac{\partial g}{\partial
d}\Bigr]\tilde d({\bf R}),
\end{equation}
where  $\tilde d ({\bf R}) = \tilde z_1 ({\bf R}) - \tilde z_2 ({\bf R})$. From
(\ref{23}) we obtain

\begin{equation}\label{30}
K_{\bf k} = - \frac{ n_{s0}}{M_H}\Bigl[\frac{1}{g n_{s0}}\Bigl(\frac{\partial E_0}{\partial d} +
n_{s0}\frac{\partial g}{\partial d}\Bigr) + \frac{1}{M_H}\frac{\partial M_H}{\partial d} \Bigr]\tilde d_{\bf k}.
\end{equation}
The expression for ${\bf E}_{int \,{\bf k}} $ directly follows from (\ref{21}):

\begin{equation}\label{31}
{\bf E}_{int \,{\bf k}} =  {e \over 2 \varepsilon l_H } F(d) i{\bf k}(z_{1{\bf k}}+z_{2{\bf k}}).
\end{equation}

After substitution of (\ref{30}) and (\ref{31}) into $\langle{\bf j}_s\rangle$ integration over angles of the
vector ${\bf k}$ reduces the last term in (\ref{27}) to zero. The remaining terms yield

\begin{equation}\label{32}
\langle{\bf j}_s\rangle =  K_{0}  \Bigl\{1 -  {1\over 2  }\Bigl[\frac{1}{g n_{s0}}\Bigl(\frac{\partial
E_0}{\partial d} + n_{s0}\frac{\partial g}{\partial d}\Bigr) + \frac{1}{M_H}\frac{\partial M_H}{\partial d}
\Bigr]^2 \langle {\tilde d}^{\,2}\rangle \Bigr\}{\bf p}_{s0}.
\end{equation}
Thus, variations of the interlayer distance lead to a decrease of the average
current. It follows from (\ref{32}) that in a bilayer system with curved
conducting layers the effective superfluid density is

\begin{equation}\label{33}
n_{ef} = n_{s0}  \Bigl\{1 -  {1\over 2  }\Bigl[\frac{1}{g n_{s0}}\Bigl(\frac{\partial E_0}{\partial d} +
n_{s0}\frac{\partial g}{\partial d}\Bigr) + \frac{1}{M_H}\frac{\partial M_H}{\partial d} \Bigr]^2 \langle
{\tilde d}^{\,2}\rangle \Bigr\}.
\end{equation}
At $d \ll l_H$ this expression is significantly simplified:

\begin{equation}\label{34}
n_{ef} = n_{s0}  \Bigl(1 -  {1\over \pi \nu_1^2 }\frac{l_H^2  \langle {\tilde d}^{\,2}\rangle}{d^4}\Bigr).
\end{equation}

Comparison of the expressions (\ref{18}) and (\ref{34}) gives that variations
of the interlayer distance suppress the superfluid density stronger than
ionized donors, if

\begin{equation}\label{35}
\frac{ \langle {\tilde d}^{\,2}\rangle}{d^2} > {1\over 4 }\frac{l_H^2 }{Z_d^2}.
\end{equation}
If we take typical values $Z_d$ = 300 nm, $l_H$ = 30 nm, $d$ = 20 nm, we find $ \langle {\tilde d}^{\,2}\rangle
^{1/2}> 1$ nm. Thus, in the case $d \lesssim l_H$  we can assume that the influence of interlayer distance
fluctuations prevails over the influence of ionized donors. Being based on this estimation, in the next section
we will consider the suppression of the superfluid transition temperature caused by variations of $ {\tilde d}$.
Note that within the approach developed here simultaneous consideration of influences of ionized donors and
interlayer distance fluctuations in $T_c$ is also possible, although it leads to quite cumbersome expressions.

 %\pagebreak

 \section{Influence of interlayer distance fluctuations on the superfluid transition temperature}

It is well known that in a 2D system of neutral bosons superfluidity is destroyed when formation of vortices
leads to decrease of the free energy (Berezinskii -- Kosterlitz -- Thouless transition) \cite{19,20}. The
transition temperature $T_c$ is found from the equation

\begin{equation}\label{36}
T_c = \frac{ \pi \hbar^2}{2 k_B} K(T_c).
\end{equation}

Since $e-h$ pairs at low density behave as a gas of Bose particles, the equality (\ref{36}) also determines the
superfluid transition temperature in the bilayer systems considered here. It is important to note that the value
of $K$ decreases both due to collective excitations of the condensate (phonons) and under the action of
inhomogeneities. Calculation of the influence of phonons does not cause difficulties (see details e. g. in
\cite{21}), and the influence of $\tilde d({\bf R})$ has been considered in the previous section. Combining both
contributions, we obtain

\begin{equation}\label{37}
K(T) = K_0 - \frac{3 \zeta (3) k_B^3 T^3 }{2\pi\hbar^2 g^2 n_{s0}} - {1\over 2 K_0 } \Bigl(\frac{\partial
K}{\partial d }\Bigr)^2\langle {\tilde d}^{\,2}\rangle,
\end{equation}
where $\zeta (3)$ = 1,202 and

\begin{equation}\label{38}
\frac{\partial K}{\partial d } =  -  K_0 \Bigl[\frac{1}{g n_{s0}}\Bigl(\frac{\partial E_0}{\partial d} +
n_{s0}\frac{\partial g}{\partial d}\Bigr) + \frac{1}{M_H}\frac{\partial M_H}{\partial d} \Bigr].
\end{equation}

Substituting $K(T)$ into (\ref{36}) yields a cubic equation for the critical temperature:

\begin{equation}\label{39}
T_c = \frac{3 \zeta (3) k_B^3 }{4 g^2 n_{s0}^2} (T_0^3 - T_c^3).
\end{equation}
Here

\begin{equation}\label{40}
T_0^3 = \frac{2\pi\hbar^2 g^2 n_{s0}^2 }{3 \zeta (3) k_B^3} \Bigl[K_0 -{1\over 2 K_0 } \Bigl(\frac{\partial
K}{\partial d }\Bigr)^2\langle {\tilde d}^{\,2}\rangle\Bigr].
\end{equation}
If the following notation is used:

\begin{equation}\label{41}
T_* = \frac{3\pi\hbar^2 }{2 k_B } \Bigl[K_0 -{1\over 2 K_0 } \Bigl(\frac{\partial K}{\partial d }\Bigr)^2\langle
{\tilde d}^{\,2}\rangle\Bigr],
\end{equation}
the critical temperature $T_c$ can be represented by a compact formula

\begin{equation}\label{42}
T_c = T_0  \Bigl\{\root 3 \of {\frac{1}{2}+\Bigl[\frac{1}{4}+\Bigl(\frac{T_0}{T_*}\Bigr)^3\Bigr]^{1/2}} + \root
3 \of {\frac{1}{2}-\Bigl[\frac{1}{4}+\Bigl(\frac{T_0}{T_*}\Bigr)^3\Bigr]^{1/2}}\Bigr\}.
\end{equation}

It is convenient to analyze the results in the graphic form. First we assume ${\tilde d}=0$ and find the
dependence $T_c(H,d)$ for a homogeneous bilayer system (Fig. \ref{f2}). It follows from Fig. 2 that at a given
value of the magnetic field the critical temperature as a function of the interlayer distance has the form of a
curve with a maximum. The maximum value of $T_c$ is reached at $d = d_{max}$, and the dependence $d_{max}(H)$
can be approximated quite well by the curve $ d_{max} = 0,4 l_H$. In the $d < d_{max}$ region the critical
temperature rapidly decreases due to strengthening  thermal fluctuations with decreasing the interaction
constant $g$. A smooth decrease of the critical temperature at $d > d_{max}$ is caused by increase of the
effective mass of the $e-h$ pair as $d$ increases. Note that this result describes the behavior of $T_c(d)$ only
qualitatively, because at large $d$ destruction of phase coherence is first of all caused by quantum
fluctuations \cite{11}, i. e. due to an effect which is beyond the self-consistent field approximation used by
us.

\begin{figure}
  
  \includegraphics[width=9 cm]{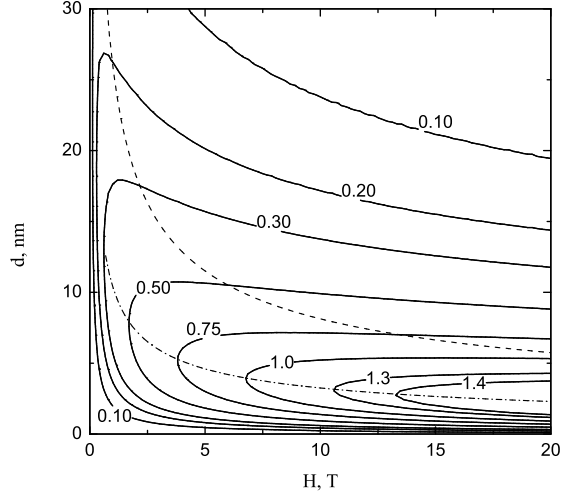}
  \caption{Temperature of the superfluid transition $T_c$ as a function of magnetic field $H$ and interlayer
distance $d$. The temperature $T_c$ is calculated for a homogeneous system at fixed filling factor $\nu_{1} =
0.1$ and dielectric constant $\varepsilon = 12.5$. The domain of applicability of the result is located to the
left and lower than the dashed line given by equality $d=l_H$; the dash-dot line $d = 0.4 l_H$ approximates the
positions of maxima of the dependence $T_c(d)$ at a given magnetic field (see main text).}\label{f2}
\end{figure}
Now we proceed to consideration of bilayer systems with random variations of the interlayer distance $ {\tilde
d} \ne 0$. The influence of such variations on the superfluid transition temperature is shown in Fig. 3, where
the curves of $T_c(d)$ are shown at different values of the mean square fluctuation $ \langle {\tilde
d}^{\,2}\rangle ^{1/2}$. It is obvious that random inhomogeneities of the interlayer distance suppress the
critical temperature. This suppression has such a character that the form of the $T_c(d)$ curves is unchanged,
but the height of the maximum of $T_c(d)$ significantly decreases and its position shifts to the range of larger
values of $d$. While analyzing Fig. \ref{f3}, of course, one must take into account that the behavior of $T_c$
in the range, where the corrections to the critical temperature are of the order of $T_c$ itself, must be
realized only as qualitatively correct.

\begin{figure}
  \includegraphics[width=9 cm]{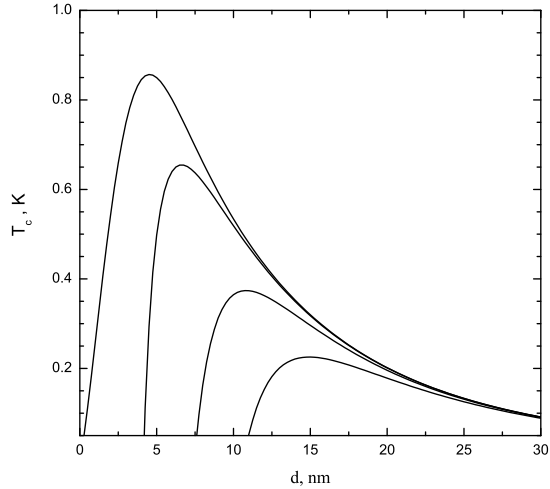}
  \caption{Influence of the interlayer distance  on the superfluid transition temperature. Curves $T_c(d)$
are calculated for a magnetic field $H=5$ T and mean square deviations $ \langle {\tilde d}^{\,2}\rangle
^{1/2}=$ 0, 0.3, 1 and 2 nm (top-down).}\label{f2}
\end{figure}

If we consider each curve in Fig. 3 as a boundary between superfluid and normal phases at corresponding $
\langle {\tilde d}^{\,2}\rangle ^{1/2}$, we can make a conclusion that at given values of temperature and
magnetic field the superfluid phase exists in a particular range of interlayer distances, $d_{c1}<d<d_{c2}$, and
the size of this interval decreases with increasing the mean square fluctuation $ \langle {\tilde
d}^{\,2}\rangle ^{1/2}$. It is obvious from Fig. 3 that the decrease of the superfluidity existence range is
mainly caused by strong influence of inhomogeneities on $d_{c1} $.

% \pagebreak

 \section{Summary}

The influence of the randomly distributed inhomogeneities on the superfluidity of the $e-h$ pairs in bilayer systems has been
considered in this article. An important and partly unexpected result is a high sensitivity of systems with small
interlayer distances ($ d \ll l_H$) to inhomogeneities. Formally, this is expressed through the appearance of the
factors $ (d/ l_H)^2$ in expressions (\ref{18}) and (\ref{34}) which describe the decrease
of the effective superfluid density $n_{ef}$ due to the electrostatic field of the ionized donors and the roughening of the
conducting layers respectively. The physical origin of the strong suppression of $n_{ef}$ is the weakness of the
interaction between the $e-h$ pairs for $ d \ll l_H$ (see Eq.~(\ref{17})).

The results of this article have been obtained under assumption that the filling
factor of the ground Landau level $\nu_{1}\ll 1 $. From our point of view this
limitation is not significant, since according to equalities (\ref{40}) and
(\ref{41}) the increase of the pair density $n_{s0}$ (i. e. the increase of $\nu_{1}
$) leads to the increase of the critical temperature $T_c$, but does not change the
picture represented in Fig. 3. Thus, one can expect that these results
qualitatively correctly describe the behavior of $T_c(d)$ in the inhomogeneous
systems with filling factors $\nu_{1}\approx 1/2 $ which are commonly used in
experiments on finding the superfluidity of the $e-h$ pairs.

Also we note that several important conclusions for future experiments can be drawn. Firstly, one of the
critical conditions for observing superfluidity in bilayer electron-hole systems is a high degree of homogeneity
of these systems. Secondly, in homogeneous systems the maximum value of $T_c$ is reached for the interlayer
distance $ d \approx 0.4 l_H$. The superfluid phase of the interwell excitons has to be looked for in the vicinity
of this value. Additionally one has to keep in mind that in the real, i. e. inhomogeneous, systems the range
of the interlayer distances $d$ at which the observation of the superfluid phase is most probable, is shifted
towards higher values (see Fig. 3).

%\pagebreak

\end{document}